\documentclass[12pt,preprint]{aastex}

\newcommand{\bdv}[1]{\mbox{\boldmath$#1$}}

\def\au{{\rm AU}} 
 
\def\kms{{\rm km}\,{\rm s}^{-1}}
\def\masyr{{\rm mas}\,{\rm yr}^{-1}}
\def\kpc{{\rm kpc}}

\def\max{{\rm max}}
\def\rel{{\rm rel}}

\def\eff{{\rm eff}}
\def\hel{{\rm hel}}
\def\geo{{\rm geo}}
\def\e{{\rm E}}
\def\bpi{{\bdv\pi}}
\def\bmu{{\bdv\mu}}

\def\bgamma{{\bdv\gamma}}

\def\bv{{\bf v}}
\def\naive{{\rm naive}}
\begin{document}
\title{{\it Spitzer} as Microlens Parallax Satellite: Mass Measurement for
the OGLE-2014-BLG-0124L Planet and its Host Star}

\author
{A.~Udalski$^{1}$,
J.~C.~Yee$^{2}$,
A.~Gould$^{3}$, 
S.~Carey$^{4}$,
W.~Zhu$^{3}$,
J.~Skowron$^{1}$,
S.~Koz{\l}owski$^{1}$,
R.~Poleski$^{1,3}$,
P.~Pietrukowicz$^{1}$,
G. Pietrzy{\'n}ski$^{1,5}$
M.K.~Szyma{\'n}ski$^{1}$,
P.~Mr{\'o}z$^{1}$,
I.~Soszy{\'n}ski$^{1}$,
K.~Ulaczyk$^{1}$,
{\L}.~Wyrzykowski$^{1,6}$,
C.~Han$^{7}$,
S.~Calchi Novati$^{8,9,10}$,
R.W.~Pogge$^{3}$\\
\normalsize{$^{1}$Warsaw University Observatory, Al.~Ujazdowskie~4, 00-478~Warszawa, Poland} \\ 
\normalsize{$^{2}$Harvard-Smithsonian Center for Astrophysics, 60 Garden St., Cambridge, MA 02138, USA} \\
\normalsize{$^{3}$Department of Astronomy, Ohio State University, 140 W. 18th Ave., Columbus, OH  43210, USA} \\
\normalsize{$^{4}$Spitzer Science Center, MS 220-6, California Institute of Technology,Pasadena, CA, USA} \\
\normalsize{$^{5}$Universidad de Concepci{\'o}n, Departamento de Astronomia, Casilla 160--C, Concepci{\'o}n, Chile} \\
\normalsize{$^{6}$Institute of Astronomy, University of Cambridge, Madingley Road, Cambridge CB3 0HA, UK} \\
\normalsize{$^{7}$Department of Physics, Chungbuk National University, Cheongju 371-763, Republic of Korea} \\
\normalsize{$^{8}$NASA Exoplanet Science Institute, MS 100-22, California Institute of Technology, Pasadena, CA 91125, USA\footnote{Sagan visiting fellow}} \\
\normalsize{$^{9}$Dipartimento di Fisica ``E. R. Caianiello'', Universit\`a di Salerno, Via Giovanni Paolo II, 84084 Fisciano (SA),\ Italy} \\
\normalsize{$^{10}$Istituto Internazionale per gli Alti Studi Scientifici (IIASS), Via G. Pellegrino 19, 84019 Vietri Sul Mare (SA), Italy}
}


\begin{abstract}
We combine {\it Spitzer} and ground-based observations to measure the
microlens parallax vector $\bpi_\e$,  and so the mass and distance of
OGLE-2014-BLG-0124L, making it the first microlensing planetary system
with a space-based parallax measurement.  The planet and star have
masses $m \sim 0.5\,M_{\rm jup}$ and $M\sim 0.7\,M_\odot$ and are
separated by $a_\perp\sim 3.1\,\au$ in projection.  The main source of
uncertainty in all these numbers (approximately 30\%, 30\%, and 20\%) is
the relatively poor measurement of the Einstein radius $\theta_\e$,
rather than uncertainty in $\pi_\e$,  which is measured with 2.5\%
precision, which compares to 22\% based on OGLE data alone.
The {\it Spitzer} data therefore provide not only a substantial
improvement in the precision of the $\pi_\e$ measurement but also the
first independent test of a ground-based $\bpi_\e$ measurement.

\end{abstract}


\section{{Introduction}
\label{sec:intro}}

Observing microlensing events from a ``parallax satellite'' is a
powerful way to constrain or measure the lens mass, as first suggested a
half century ago by \citet{refsdal66}.  This idea has acquired increased
importance as microlensing planet searches have gained momentum, since
obtaining masses and distances for these systems is the biggest
challenge facing the microlensing technique.  By chance, the typical
scale of Galactic microlensing events is ${\cal O}(\au)$, which is why
it is a good method to find extrasolar planets \citep{gouldloeb}.  By
the same token, a microlensing satellite must be in solar orbit in order
that its parallax observations (combined with those from Earth) probe
this distance scale. Hence, it was long recognized that the {\it
Spitzer} spacecraft, using the $3.6\,\mu$m channel on its IRAC camera
would make an excellent microlensing parallax satellite \citep{gould99}.

Nevertheless, until this year, {\it Spitzer} had  made only one
microlensing parallax measurement, which was of an event with a
serendipitously bright source star in the Small Magellanic Cloud
\citep{dong07}.  In 2014, however, we received 100 hours of observing
time to carry out a pilot program of microlens parallax observations
toward the  Galactic bulge, with the primary aim of characterizing
planetary events.

Here, we report on the first result from that program, a mass and
distance measurement for the planet OGLE-2014-BLG-0124Lb.


The microlens parallax $\bpi_\e$ is a two dimensional vector defined by
\begin{equation}
\bpi_\e \equiv {\pi_\rel\over \theta_\e}\,{\bmu\over\mu}.
\label{eqn:piedef}
\end{equation}
The magnitude of this vector is the lens-source relative parallax
$\pi_\rel$ ($\pi_\rel = \pi_L - \pi_S$) scaled to the Einstein radius
$\theta_\e$.  This is because $\pi_\rel$ determines how much the lens
and source will be displaced in angular separation as the observer
changes location, while $\theta_\e$ sets the angular scale of
microlensing phenomena, i.e., the mapping of the physical effect of the
displacement onto the lightcurve.  The direction of $\bpi_\e$ is the
same as that of the lens-source relative proper motion $\bmu$ because 
this direction determines how the lens-source displacement will evolve
with time.  See Figure 1 of \citet{gouldhorne} for a didactic
explanation.

Combining Equation~(\ref{eqn:piedef}) with the definition of $\theta_\e$,
\begin{equation}
\theta_\e \equiv \sqrt{\kappa M \pi_\rel};
\quad
\kappa \equiv {4 G \over c^2\au}\simeq 8.14\,{{\rm mas}\over M_\odot},
\label{eqn:thetae}
\end{equation}
yields a solution for the lens mass $M$ 
\begin{equation}
M = {\theta_\e\over\kappa \pi_\e} = {\mu t_\e\over\kappa \pi_\e}.
\label{eqn:meqn}
\end{equation}
Hence, if $\pi_\e$ and $\theta_\e$ are both measured, the mass is
determined from the first form of this equation.  However, even if
$\theta_\e$ is not measured, the second form of
Equation~(\ref{eqn:meqn}) gives a good estimate of the mass because
$t_\e$ is almost always known quite well and the great majority of
microlensing events will have proper motions within a factor of 2 of
$\mu\sim 4\,\masyr$.  By contrast, if neither $\theta_\e$ nor $\pi_\e$
is measured, a mass estimate based on $t_\e$ alone is extremely crude. 
See Figure~1 from \citet{gould00}.

Since $\pi_\e^2 = \pi_\rel/\kappa M$, typical values are $\pi_\e\sim
0.3$ for lenses in the Galactic disk and $\pi_\e\sim 0.03$ for lenses in
the Galactic bulge.  Hence, the projected Einstein radius $\tilde r_\e
\equiv \au/\pi_\e$ typically lies in  the range from one $\au$ to several tens
of $\au$.  Thus, to see a substantially different event from that seen
from Earth requires that the satellite be in solar orbit.

\section{{Observations}
\label{sec:obs}}

We combine observations from two observatories, {\it Spitzer}
and the Optical Gravitational Lensing Experiment (OGLE).  

{\subsection{Spitzer Program}
\label{sec:spitzprog}}

The {\it Spitzer} observations were carried out under a 100 hour pilot
program granted by the Director to determine the feasibility of {\it
Spitzer} microlens parallax observations toward the Galactic bulge. Due
to Sun-angle viewing constraints, targets near the ecliptic (including
bulge microlensing fields) are observable for two $\sim 38\,$day
continuous viewing periods per 372 day orbital period. Our observation
period (2014 June 6 to July 12) was chosen to maximize observability of
likely targets, which are grouped in a relatively narrow range of Right
Ascension near 18.0 hours\footnote{Note that although these targets are
equally visible from {\it Spitzer} during an interval that is 186 days
later, they would be behind the Sun as viewed from  Earth, making
parallax measurements impossible.}. Targets were observed during 38
2.63-hr epochs, separated by roughly one day, from  HJD$^\prime=$ HJD -
2450000 = 6814.0 to 6850.0.

Each observation consisted of 6 dithered 30s exposures in a fixed
pattern using the 3.6$\,\mu$m channel on IRAC.  Taking account of
various overheads, including time to slew to new targets, this permitted
observation of about 34 targets per epoch\footnote{Note that the slew
time for this program is significantly shorter than is typical for {\it
Spitzer} because the targets are grouped with a few degrees of each
other on the sky.}.

The process of choosing targets and the cadence at which they were
observed was complex.  A special observing mode was developed
specifically for this project.  The 38 2.6-hr epochs were set aside in
the {\it Spitzer} schedule well in advance.  Then, each Monday at UT
15:00, draft sequences were uploaded to {\it Spitzer} operations for
observations to be carried out Thursday to Wednesday (with some slight
variations).  These sequences were then vetted for suitability,
primarily Sun-angle constraints, and then uploaded to the spacecraft.

Thus, the first problem was to identify targets that could usefully be
observed 3 to 9 days in advance of the actual observations.  The first
reason that this is challenging is that it is usually difficult to
predict the evolution of a microlensing lightcurve from the rising wing,
particularly at times well before peak.  The characteristic timescale of
microlensing events is $t_\e\sim 25\,$days.  Hence, for example, 9 days
before peak an event with a typical source magnitude $I_s=19$ would be
only 1 mag brighter, $I=18$, meaning that ground-based photometry would
be relatively poor, allowing only a crude prediction of its evolution. 
Such predictions are typically consistent with  a broad range of fits,
extending from the event peaking not much brighter than its current
brightness (implying it would be unobservably faint in {\it Spitzer}
data) to peaking at very high magnification (which would allow an
unambiguous {\it Spitzer} parallax measurement).   The second reason is
that {\it Spitzer} will necessarily see a different lightcurve than the
one from the ground (this is the point of parallax observations!). 
Observations are much more likely to yield good parallax measurements if
the event peaks as seen by {\it Spitzer}, but depending on the value of
the parallax, this peak could be very similar to the peak time seen from
Earth or days or weeks earlier or later.

To address the first challenge JCY wrote software to automatically fit
all ongoing microlensing events and assess whether or not they met
criteria for inclusion in the {\it Spitzer} observation campaign. This
software was tested on OGLE data from the 2013 microlensing season and
used to simulate the {\it Spitzer} observations by fitting the data for each
event up to a certain cutoff date, and repeating for successive weeks.
Then JCY and AG estimated the correctness of these automated choices by
comparing to fits of the complete light curves, that is determining
whether or not an event classification based on incomplete data was
correct when compared to the final, known properties of that event. This
served as the basis both for fine-tuning the software and for learning
when to manually override it. These lessons were then applied each week
by JCY+AG to the actual choice of targets.

To further expedite this process, OGLE set up a special real-time
reduction pipeline for potential targets under consideration, with
updates lagging observations by just a few minutes.  This permitted
robust construction of a trial protocol at about UT 03:00 Monday, and
late-time tweaking based on the most recent OGLE data (typically ending
at UT 10:00) for final internal vetting and translation into a set of
``Astronomical Observation Requests'' before uploading to {\it Spitzer}
operations.

The next problem was to determine the cadence.  The program limited
observations to 2.6 hour windows roughly once per day.  This precluded
using {\it Spitzer} to find planets, since this requires observations at
several-to-many times per day.  Moreover, since there were usually more
than 34 targets that could usefully be observed during a given week, not
all of the targets could be observed at every epoch. Targets were thus
divided into ``daily'', ``moderate'', and ``low'' cadence.  The first
were observed every epoch, the second were observed most epochs, and
the third were observed about 1/3 to 1/2 of the epochs.  In addition, a
few targets were regarded as ``very high priority'' and so were slated
to be observed more than once per epoch. Particularly during the first
week, when there were many targets that had just peaked (and of course
had not yet been observed), targets that were predicted for peak many
weeks in the future were downgraded in priority.  This constraint
directly impacted observations of OGLE-2014-BLG-0124.

{\subsection{OGLE Observations}
\label{sec:ogleobs}}

On February 22, 2014 OGLE alerted the community to a new microlensing
event OGLE-2014-BLG-0124 based on observations with the 1.4 deg$^2$
camera on its  1.3m Warsaw Telescope at the Las Campanas Observatory in
Chile using its Early Warning System (EWS) real-time event detection
software \citep{ews1,ews2}.  Most observations were in $I$-band, with a
total of 20 $V$-band observations during 2014 to determine the source
color. The source star lies at (RA, Dec) = (18:02:29.21, $-28$:23:46.5)
in OGLE field BLG512, which is observed at OGLE's highest cadence, about
once every 20 minutes.

On June 29 UT 17:05, our group alerted the microlensing community to an
anomaly in this event, at that time of unknown nature, based on analysis
of OGLE data from the special pipeline described above.  While in some
cases (e.g., \citealt{mb11293}) OGLE responds to such alerts by
increasing its cadence, it did not do so in this case because of the
high cadence already assigned to this field.  Hence, OGLE observations
are exactly what they would have been if the anomaly had not been
noticed.

For the final analysis the OGLE dataset was re-reduced. Optimal
photometry was derived with the standard OGLE photometric pipeline
\citep{ews2} tuned-up to the OGLE-IV observing set-up, after deriving
accurate centroid of the source star.  

\subsection{Spitzer Cadence}

At the decision time (June 2 UT 15:00, HJD$^\prime$ 6811.1) for the
first week of {\it Spitzer} observations, OGLE-2014-BLG-0124 was
regarded as a promising target, but because it appeared to be peaking
30--40 days in the future, it was assigned ``moderate'' priority, which
because of the large number of targets in the first week implied that it
was observed in only three of the first eight epochs.  The following
week, it was degraded to ``low'' priority because its estimated peak
receded roughly one week into the future. Nevertheless, because  the
total number of targets fell from 44 to 37, OGLE-2014-BLG-0124 was
observed during four of the six epochs scheduled that week.  Since the
peak was approaching, it was raised back to ``moderate'' priority in the
third week and observed in six out of eight epochs, and then to
``daily'' priority in the fourth week and observed in all seven epochs. 
It was the review of events in preparation for the fifth week that led
to the recognition that OGLE-2014-BLG-0124 was undergoing an anomaly
(Section~\ref{sec:ogleobs}), and hence it was placed at top priority. 
In addition, as the week proceeded, the events lying toward the west of
the microlensing field gradually moved beyond the allowed Sun-angle
range, which permitted more observations of those (like
OGLE-2014-BLG-0124) that lay relatively to the East. As a result, it was
observed a total of 20 times in eight epochs.

\begin{figure}
\plotone{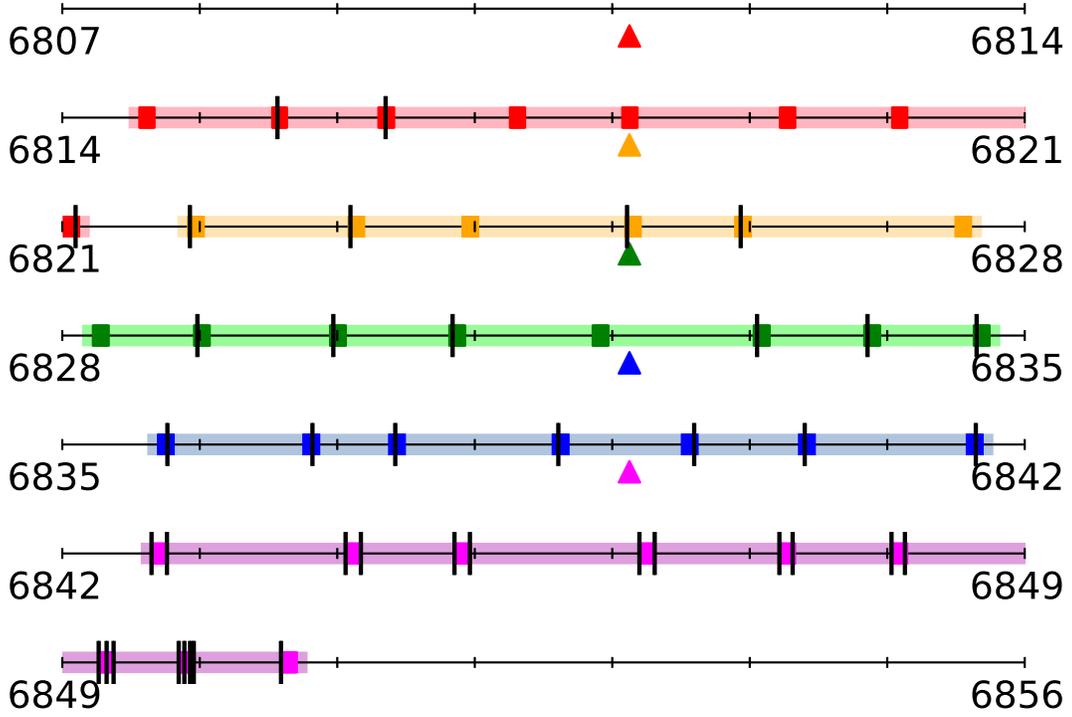}
\caption{Timeline (HJD$^\prime$) of Spitzer observations. Each Spitzer ``week'' 
of observations is color-coded. The triangle indicates the date the
microlensing targets were submitted to the Spitzer Science
Center for observations during the corresponding ``week''
indicated by the light bands. The solid sections of the bands
indicate the blocks allocated to microlensing observations,
which were taken approximately once per day. The black, vertical
lines indicate the specific observations of OGLE-2014-BLG-0124.
These observations were more sparse early in the Spitzer
campaign and became more dense as the event neared peak 
(as seen from Earth) and was discovered to host a planet.
}
\label{fig:timeline}
\end{figure}

In fact, due to the particular configuration of the event, the most
crucial observations turned out to be those during the first 10 days
when the event was rated as ``low'' to ``moderate'' priority.  See
Figures~\ref{fig:timeline} and \ref{fig:lc}.

The {\it Spitzer} data were reduced using DoPhot \citep{dophot}
after experimentation with several software packages.  DoPhot's superior
performance may be related to the fact that the OGLE-2014-BLG-0124
source is isolated on scales of the {\it Spitzer} point spread function (PSF),
but this is a matter of ongoing investigation as we continue to analyze
events from the {\it Spitzer} microlens program.

\section{{Heuristic Analysis}
\label{sec:heuristic}}

\begin{figure}
\plotone{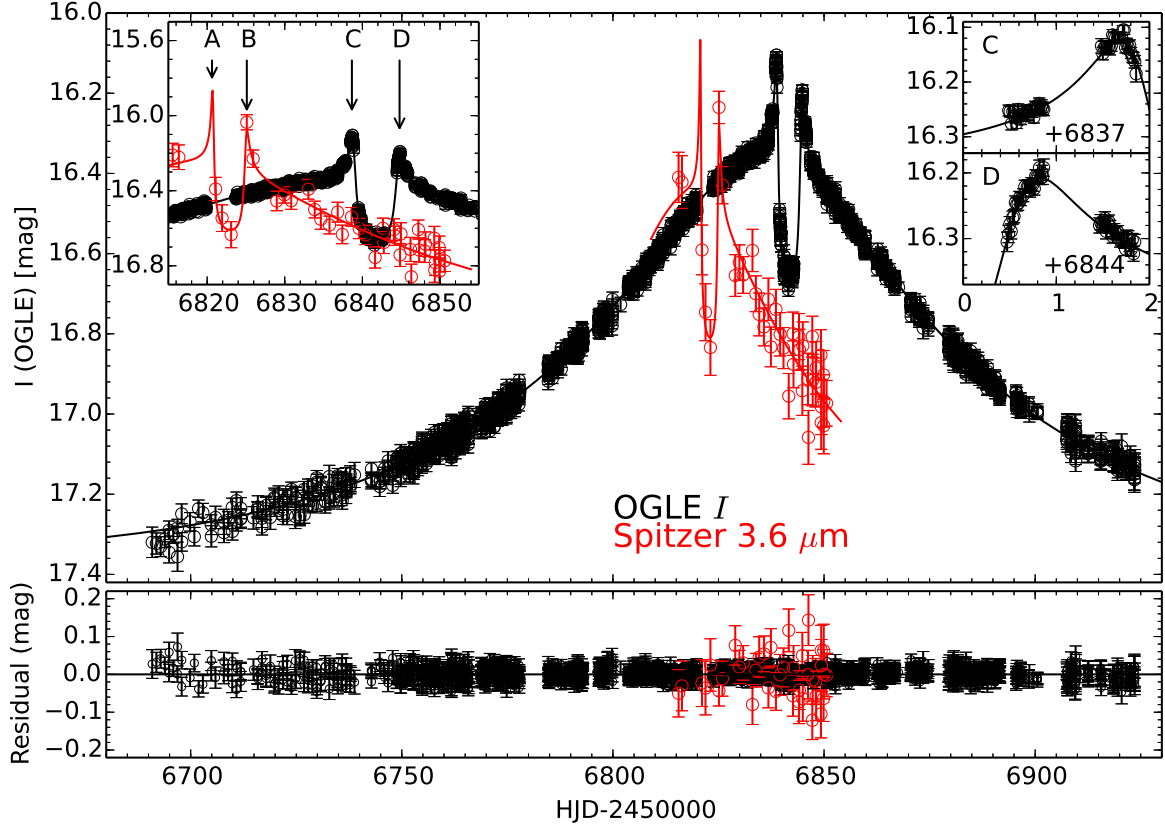}
\caption{Lightcurve and residuals for planetary model of OGLE-2014-BLG-0124
as observed from Earth by OGLE in $I$ band (black) and by {\it Spitzer} 
at $3.6\,\mu$m (red), 
which was located $\sim 1\,\au$ East of Earth in projection
at the time of the observations.  Simple
inspection of the OGLE lightcurve features shows that this is Jovian planet,
while the fact that {\it Spitzer} observed similar features 20 days earlier
demonstrates that the lens is moving $\tilde v\sim 105\,\kms$ due East
projected on the plane of the sky (Section~\ref{sec:heuristic}).  
Detailed model-fitting confirms and refines this by-eye analysis
(Section~\ref{sec:phys}).  Note that in the left inset, the {\it Spitzer}
light curve is aligned to the OGLE system (as is customary), but it is
displaced by 0.2 mag in the main diagram, for clarity.
}
\label{fig:lc}
\end{figure}

\begin{figure}
\plotone{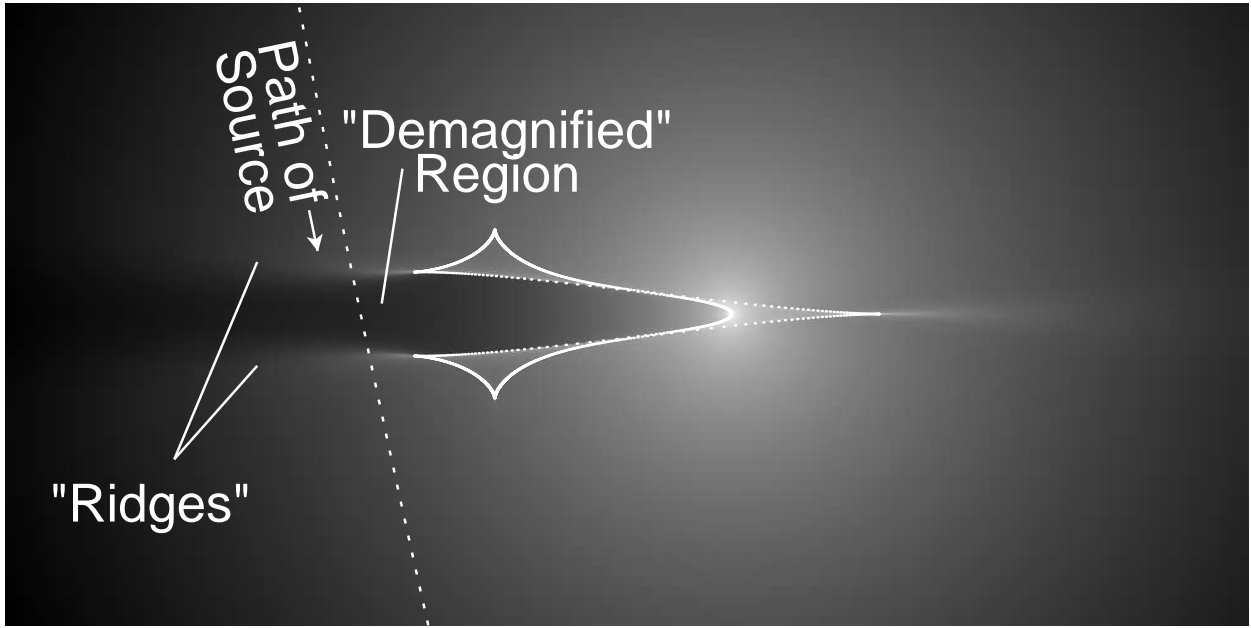}
\caption{Magnification map for caustic region of OGLE-2014-BLG-0124
in standard orientation with planet to right.
As the source passes over the ``demagnified'' region (darker tones),
the minor image due (to the primary lens) passes very close to the
planet, which is off the figure to the right.  Because the minor image
is unstable, it is easily destroyed by the planet, which accounts
for the relative demagnification.  Two triangular caustic regions
flank the deepest part of this demagnification.  The source does
not cross these causitics, but does cross the two ridges that extend
from the cusps, toward the left.  It is these ridges that are responsible
for the two bumps near $t=6820$ and $t=6825$ (from {\it Spitzer}) or
$t=6839$ and $t=6845$ (from Earth) in Figure~\ref{fig:lc}.
}
\label{fig:magmap}
\end{figure}

The most prominent feature in the OGLE lightcurve  (black points,
Figure~\ref{fig:lc}) is a strong dip very near what would otherwise be
the peak of the lightcurve (HJD$^\prime\sim 6842$).  The dip is flanked
by two peaks (highlighted in the insets),  each of which is pronounced
but neither of which displays the violent breaks characteristic of
caustic crossings. This dip must be due to an interaction between a
planet and the minor image created by the host star in the underlying
microlensing event.  That is, in the absence of a planet, the host will
break the source light into two magnified images, a major image outside
the Einstein ring on the same side as the source and a minor image
inside the Einstein ring on the opposite side from the source  (e.g.,
\citealt{gaudi12}).  Being at a saddle point of the time-delay surface,
the minor image is highly unstable to perturbations, and is virtually
annihilated if a planet lies in or very near its path.   These
(relatively) demagnified regions are always flanked by two triangular
caustics (see Figure~\ref{fig:magmap}). If the source had passed over
these caustics, it would have shown a sharp break in the lightcurve
because the magnification of a point source diverges to infinity as it
approaches a caustic.  Hence, from the form of the perturbation, it is
clear that the source passed close to these caustics but not directly
over them.  Because the two peaks are of nearly equal height, the source
passed so the angle $\alpha$ between its path and planet-star axis is
rougly $90^\circ$.

The {\it Spitzer} lightcurve (red points, Figure~\ref{fig:lc}) shows
very similar morphological features but displaced about 19.5 days
earlier in time. The velocity of the lens relative to the source
(projected onto the observer plane) $\tilde \bv$ is easily measured by
combining information from the {\it Spitzer} lightcurve and the OGLE
lightcurve. This is the most robustly measured quantity derived from the
lightcurve, and it is related to the parallax vector by
\begin{equation}
\bpi_\e = {\au\over t_\e}\,{\tilde \bv\over \tilde v^2}.
\label{eqn:vtpar}
\end{equation}

Projected on the plane of the sky, {\it Spitzer}'s position at the time
it saw the dip (HJD$^\prime$ 6822.5) was about 1.17 AU away from where
the Earth was when it saw the dip (HJD$^\prime$ 6842), basically due
West of Earth.  Hence, the projected velocity of the lens relative to the source
(in the heliocentric frame) along this direction is $\tilde v_{\hel,E}
\sim 1.17\,\au/(19.5\,{\rm day})\sim 105\,\kms$.  On the other hand the
fact that the morphology is similar shows that the source passed the
caustic structure at a similar impact parameter perpendicular to its
trajectory (i.e., in the North direction).  Hence $\tilde v_{\hel,N}
\sim 0$.  One converts from heliocentric to geocentric frames by $\tilde
\bv_\hel = \tilde \bv_\geo + \bv_{\oplus,\perp}$ where
$\bv_{\oplus,\perp}({\rm N,E}) \simeq (0,30)\,\kms$ is the velocity of
Earth projected on the sky at the peak of the event.  Hence,
\begin{equation}
\tilde \bv_\hel = \tilde \bv_\geo + \bv_{\oplus,\perp} \simeq (0,105)\,\kms.
\quad
\tilde \bv_\geo({\rm N,E})\simeq (0,75)\,\kms;
\label{eqn:vtilde}
\end{equation}
This result is robust and does not depend in any way on the details of
the analysis.

Next we estimate the planet-host mass ratio $q$ and the planet-host
projected separation $s$ in units of the Einstein radius making use of
three noteworthy facts.  First, because the perturbation affects the
minor image, $s<1$. Second, by making the approximation that the planet
passes directly over the minor image, we can express the position of the
source as $u = 1/s - s$.  Third, because the perturbation occurs close
to the time of the peak,  $u_{\rm perturbation}\simeq u_0$, i.e.,
$u_0\simeq 1/s - s$.

The impact parameter between the source and lens $u_0$ can be estimated
from the peak magnification of the event $A_\max$. As we show
immediately below, the source star is significantly blended with another
star or stars that lie within the PSF but that
do not participate in the event.  Nevertheless, for simplicity of
exposition we initially assume that the source is not blended and then
subsequently incorporate blending into the analysis. Under this
assumption, the fact that the peak of the underlying point-lens event is
a magnitude brighter than baseline implies a peak magnification
$A_\max^\naive=2.5$ and thus an impact parameter $u_0^\naive \sim
1/A_\max^\naive \sim 0.40$ and $s\sim 0.82$. Additionally, the fact that
the event becomes a factor 1.34 brighter (corresponding to entering
$u=1$) roughly 60 days before peak, implies $t_\e^\naive=60\,$days.

We can now make use of the analytic estimate of \citet{han06} for the
perpendicular separation $\eta_{c,-}$ (normalized to $\theta_\e$)
between the the planet-star axis and the inner edge of the triangular
planetary caustic due to a minor-image perturbation \citep[see Figure 2
of ][]{han06},
\begin{equation}
\eta_{c,-}^2 \simeq 4 {q\over s}\biggl({1\over s} - s\biggr)
\end{equation}
to estimate $q$.
Because the source passes nearly perpendicular to the planet-star axis,
we have $\eta_{c,-}\simeq \Delta t_{c,-}/t_\e$, where $\Delta t_{c,-}=3\,$days
is half the time interval between the two peaks.  Then solving for
$q$ yields
\begin{equation}
q = {s\over 4u_0}\biggl({\Delta t_{c,-}\over t_\e}\biggr)^2
= {s\over 4}{\Delta t_{c,-}^2\over t_\e t_\eff} = 
1.56\times 10^{-3}s\,
\biggl({t_\e\over 60\, \rm day}\biggr)^{-1}
\biggl({t_\eff\over 24\,\rm day}\biggr)^{-1},
\end{equation}
where $t_\eff\equiv u_0 t_\e$ is the effective timescale.  Now, whereas
$t_\e$ is very sensitive to blending (because a fainter source requires
higher magnification -- so further into the Einstein ring -- to achieve
a given increase in flux), $t_\eff$ is not.  In addition,
$u_0<u_0^\naive=0.40$ implies $s>0.82$, i.e., close to unity in any
case. Thus to first order, $q$ is inversely proportional to $t_\e$. This
implies a Jovian mass ratio unless the blended flux were many times
higher than the source flux, in which case the mass ratio would be
substantially lower.

Finally, we note that the absolute position of the source, which can be
determined very precisely on difference images because the source is
then isolated from all blends, is displaced from the naive ``baseline
object'' by 80 mas. Additionally, given that the source and blend are
not visibly separable in the best seeing images, they must be closer
than 800 mas. The combination of these facts means that the blend must
contribute at least 10\% of the light. However, precise determination of
the blending requires detailed modeling, to which we now turn.

\section{{Lightcurve Analysis}
\label{sec:anal}}

In addition to the parameters mentioned in the previous section ($u_0,\
t_\e,\ q,\ s,\ \alpha,\ \bpi_\e$) and $t_0$ (where $t=t_0$ at $u=u_0$),
we include three additional parameters in the modeling.  The first is
$\rho\equiv \theta_*/\theta_\e$ where $\theta_*$ is the angular radius
of the source star. This is closely related to the source radius
self-crossing time, $t_*\equiv \rho t_\e$.  Any sharp breaks in the
underlying magnification pattern will be smoothed out on the scale of
$t_*$, which is how it is normally measured.  In fact, there are no such
sharp breaks because there are no caustic crossings.  However, the
ridges of magnification seen in Figure~\ref{fig:magmap} that give rise
to the two bumps near the peak of the lightcurve are relatively sharp
and so may be sensitive to $\rho$.

Second, we allow for orbital motion of the planet-star system. We
consider only two-dimensional motion in the plane of the  sky, which we
parameterize by $ds/dt$ (a uniform rate of change of planet-host
separation) and $d\alpha/dt$ (a uniform rate of change in position
angle).  Because the orbital period is likely to be of order several
years while the baseline of measurement between caustic features seen in
the Earth and {\it Spitzer} lightcurves is only about 22 days, we do not
expect to have sensitivity to additional parameters. In fact, we will
see that even one of these two orbital parameters is poorly constrained
so there is no basis to include additional ones.

Thus, there are 11 model parameters 
$(t_0,u_0,t_\e,\rho,\pi_{\e,E},\pi_{\rm E,N},s,q,\alpha,ds/dt,d\alpha/dt)$,
plus two flux parameters $(f_s,f_b)$ for each observatory.  For
completeness, we specify the sign conventions for $u_0$, $\alpha$, and
$d\alpha/dt$. We designate $u_0>0$ if the moving lens passes the source
on its right. We designate $\alpha$ as the (counterclockwise) angle made
by the star-to-planet axis relative to the lens-source relative proper
motion at the fiducial time, which we choose to be $t_{0,{\rm par}}=
6842$ (see below). We designate $d\alpha/dt$ to be positive if the
projected orbit of the planet is counterclockwise.

We adopt limb darkening coefficients $u_V,u_I = (0.68,0.53)$
corresponding to $(\Gamma_V,\Gamma_I)=(0.59,0.43)$ based on the models
of \citet{claret00} and the source characterization described in
Section~\ref{sec:phys}.  For the {\it Spitzer} 3.6$\,\mu$m band we adopt
$u_{3.6}=0.22$, and so  $\Gamma_{3.6}= 0.16$, which we extrapolate from
the long-wavelength values calculated by \citet{claret00}.

As is customary, we conduct the modeling in the geocentric frame
(defined as the moving frame of Earth at $t_{0,{\rm par}}=6842$).  This
time is close to the midpoint of the two cusp-approaches observed by
OGLE (see Figure~\ref{fig:lc}), which is when the angular orientation of
the planet-host system is best defined (and so has the smallest formal
error). It may seem more natural to use the heliocentric frame, given
that we have observations from two different heliocentric platforms. 
However, we adopt the geocentric frame for two reasons.  First, this
permits the simplest comparison to results derived without {\it Spitzer}
data. Second, the geocentric computational formalism is well
established, so keeping it minimizes the chance of error.  From an
algorithmic point of view, {\it Spitzer}'s orbital motion is
incorporated as a stand-in for the usual ``terrestrial parallax'' term.
That is, whereas other observatories are displaced from Earth's center
according to their location and the sidereal time, {\it Spitzer} is
displaced from Earth's center according to its tabulated distance and
position on the sky as seen from Earth.

As usual, we use the point source approximation for epochs that are far
from the caustics and the hexadecapole approximation
\citep{pejcha09,gould08} at intermediate distances.  For epochs that are
near or on crossing caustics, we use contour integration \citep{gg97}. 
In practice, contour integration is not needed at all for the
ground-based data and is used for only 5 of the {\it Spitzer} data
points, i.e., those that might conceivably pass close to a caustic. To
accommodate limb darkening, we divide the surface into 10 annuli,
although this is severe overkill in {\it Spitzer}'s case because of its
low value of $\Gamma=0.16$.

We both search for the minimum and find the likelihood distribution of
parameter combinations using a Markov Chain Monte Carlo (MCMC).

{\subsection{Estimate of $\theta_*$}
\label{sec:thetastar}}

Before discussing the model parameters, we focus first on the flux
parameters, which enable a determination of $\theta_*$.  Based on
calibrated OGLE magnitudes, we find $f_{s,\rm ogle} = 0.579\pm 0.013$,
$f_{b,\rm ogle} = 1.213\pm 0.013$,  in a system in which $f=1$
corresponds to an $I=18$ star,  i.e., $I_s = 18.59\pm 0.02$,
$I_b=17.79\pm0.01$. Using the standard approach \citep{ob03262}, we
determine the dereddened source brightness $I_{s,0}=17.57$ from the
offset from the red clump using tabulated clump brightness as a function
of position from \citet{nataf13}.  Similarly, we determine the apparent
$(V-I)_s$ color from regression of $V$ and $I$ flux over the event
(i.e., without reference to any model), and then find $(V-I)_0=0.70$
from the offset to the clump,  with assumed intrinsic color of
$(V-I)_{0,\rm cl}=1.06$ \citep{bensby13}. We then convert from $(V-I)$
to $(V-K)$ using the empirical color-color relations of \citet{bb88} and
finally estimate the source radius using the color/surface-brightness
relation of \citet{kervella04}.  We find
\begin{equation}
\theta_* = 0.95\pm 0.07\,\mu{\rm as}.
\label{eqn:thetastarz}
\end{equation}
The error is completely dominated by the 0.05 mag error in the
derivation of the intrinsic source color \citep{bensby13}, and an
adopted 0.1 mag error for vertical centroiding of the clump.

{\subsection{Physical Constraints on Two Parameters}
\label{sec:physcont}}

We find that the fits to lightcurve data leave two parameters poorly
constrained: $\rho$ and $d\alpha/dt$.  Although both distributions are
actually well-confined, in both cases a substantial fraction of the
parameter space corresponds to unphysical solutions.  This is not in
itself worrisome: the requirement of consistency with nature only
demands that physical solutions be allowed, not that unphysical
solutions be excluded by the data.  However, it does oblige us to
outline the relation between physically allowed and excluded solutions
before suppressing the latter.

In the case of $\rho$, there is a well-defined range $0<\rho<0.0025$
that is permitted by the lightcurve data at the $3\,\sigma$ level.  The
upper bound, which corresponds to $t_*=0.38\,$days, comes about because
such a long crossing time would be inconsistent (at $3\,\sigma$) with
the curvature seen in the OGLE lightcurve over the peaks. The lower
bound is strictly enforced by the positivity of stellar radii. However,
from a pure lightcurve perspective, $\rho=0$ solutions are consistent at
the $1\,\sigma$ level.  Nevertheless, arbitrarily low values of $\rho$
are not permitted physically because the lens mass and distance can be
expressed,
\begin{equation}
M = {\theta_*/\kappa \pi_\e\over\rho} = 1.2\,M_\odot{6.5\times 10^{-4}\over\rho};
\qquad
\pi_\rel = {\pi_\e \theta_*\over \rho} = 0.21\,{\rm mas}{6.5\times 10^{-4}\over\rho}.
\label{eqn:md2}
\end{equation}
Since $\pi_\e=0.15$ is very well determined from the lightcurve fits
and, as we discussed in Section~\ref{sec:thetastar}, $\theta_*$ is also
well determined, the numerators of both expressions in
Equation~(\ref{eqn:md2}) are also well-determined.  Hence, as $\rho$
decreases, both $M$ and $\pi_\rel$ increase, i.e., the host gets closer
and more luminous, hence  brighter.  The final expressions show our
adopted limit.  That is, at $\pi_\rel=0.21\,$mas ($D_L = 3.1\,\kpc$) and even
assuming that the host star lay behind all the dust seen toward the
source ($A_I=1.06$), the absolute magnitude of the lens is constrained
to be  $M_{I,L} > I_b - A_I - 5\log(D_L/10{\rm pc}) = 4.35$ which is
considerably dimmer than any $M=1.2\,M_\odot$ star.  Note that this
limit $(\rho>6.5\times 10^{-4})$ is quite consistent with the ``best
fit'' value of $\rho\sim 10^{-3}$, although as emphasized above, this
detection of $\rho>0$ is statistically quite marginal.

Second, at the $3\sigma$ level, $d\alpha/dt$ is constrained by the
lightcurve to the range $0.5<(d\alpha/dt){\rm yr} < 5$.  However,
sufficiently large values of $d\alpha/dt$ lead to unbound systems.  This
is quantified by the ratio of projected kinetic to potential energy
\citep{dong09},
\begin{equation}
\beta\equiv\biggl({E_{\rm kin}\over E_{\rm pot}}\biggr)_\perp
= {\kappa M_\odot ({\rm yr})^2
\over 8\pi^2}{\pi_\e s^3\gamma^2\over
\theta_\e(\pi_\e + \pi_s/\theta_\e)^3},
\label{eqn:betadef}
\end{equation}
where 
\begin{equation}
\bgamma = (\gamma_\parallel,\gamma_\perp)\equiv 
\biggl({ds/dt\over s},{d\alpha\over dt}\biggr)
\label{eqn:gammadef}.
\end{equation}
In this case, one cannot write the resulting limit in such a simple form
as was the case for $\rho$. However, adopting a typical value
$\rho=10^{-3}$ (and therefore  $\theta_\e=0.95\,$mas) for illustration,
and noting that $\gamma_\parallel$  is constrained to a range that
renders it irrelevant to this calculation, we obtain
$\beta=0.59[\gamma_\perp{\rm yr}]^2$.  Since $\beta>1$ implies an
unbound system, values of $|\gamma_\perp|>1.3\,{\rm yr}^{-1}$ are
forbidden. This evaluation strictly applies only for solutions with
$\rho=10^{-3}$, but actually it evolves only slowly over the allowed
range $0.65<10^3\rho<2.5$.  Typical values expected for $\beta$ are
0.2--0.6, which occur for $|\gamma_\perp|\sim 1\,{\rm yr}^{-1}$.  While
the best-fit value is  $\gamma_\perp=2\,{\rm yr}^{-1}$, values of
$\gamma_\perp\sim 1\,{\rm yr}^{-1}$ are disfavored at only
$1.5\,\sigma$.  Hence, we conclude that physically allowed systems are
close to the overall $\chi^2$ minimum and therefore we are justified in
imposing physical constraints to obtain our final solution.

\section{{Physical Parameters}
\label{sec:phys}}

Following the arguments in Section~\ref{sec:physcont}, we impose the
following two constraints,
\begin{equation}
M<1.2\,M_\odot,
\qquad
\beta<1,
\label{eqn:constraints}
\end{equation}
on output chains from our MCMC to obtain final parameters 
(Table~\ref{tab:ulens_so}) and physical parameters
(Table~\ref{tab:phys_so}).   We also considered using the more tapered
prior on $\beta$ introduced by \citet{ob120406}. However, this did not
have a perceptible effect on either the values or  the errors reported
in Tables~\ref{tab:ulens_so} and \ref{tab:phys_so}. Therefore, we
adopted the more conservative constraint in
Equation~(\ref{eqn:constraints}).

Our solution indicates a $0.5\,M_{\rm jup}$  planet orbiting a
$0.7\,M_\odot$ star that is 4.1 kpc from the Sun, with a projected
separation of 3.1 AU. This is very close to being a scaled down version
of our own Jupiter, with host mass, planet mass, and physical separation
(estimated as $\sqrt{3/2}$ larger than projected separation) all reduced
by a factor $\sim 0.6$.

{\subsection{Discrete Degeneracies}
\label{sec:discrete}}

In his original paper, \citet{refsdal66} already noted that space-based
parallaxes for point-lens events are subject to a four-fold discrete
degeneracy. This is because the satellite and Earth observatories each
see two ``bumps'',  each with different $t_0$ and $u_0$, and the
parallax is effectively reconstructed from the differences in these
quantities
\begin{equation}
\bpi_\e = {\au\over D_\perp}\biggl({\Delta t_0\over t_\e},\Delta u_0\biggr),
\label{eqn:sbpar}
\end{equation}
where, $\Delta t_0=t_{0,\rm sat}-t_{0,\oplus}$, $\Delta u_0=u_{0,\rm
sat}-u_{0,\oplus}$, and ${\bf D}_\perp$ is the projected separation
vector of the Earth and satellite, whose direction sets the orientation
of the $\bpi_\e$ coordinate system. However, whereas $\Delta t_0$ is
unambiguously determined from this procedure, $u_0$ is actually a signed
quantity whose amplitude is recovered from simple point-lens events but
whose sign is not.  Hence, there are two solutions $\Delta u_{0,-,\pm} =
\pm(|u_{0,\rm sat}|-|u_{0,\oplus})|)$ for which the satellite and Earth
observe the source trajectory on the same side of the lens as each other
(with the ``$\pm$'' designating which side this is), and two others  
$\Delta u_{0,+,\pm} = \pm(|u_{0,\rm sat}|+|u_{0,\oplus})|)$ for which
the source trajectories are seen on opposite sides of the lens
\citep{gould94}.

The first of these degeneracies is actually an extension to space-based
parallaxes of the $\pm u_0$ ``constant acceleration degeneracy'' for
ground-based parallaxes discovered almost 40 year later by 
\citet{smith03}, and which is extended to binary lenses by
\citet{ob09020}. This degeneracy results in a different {\it direction}
of the parallax vector.

The second degeneracy is much more important than the first because it
leads to a different {\it amplitude} of the parallax vector, rather than
just a different direction.  That is, the amplitude of $\bpi_\e$ in
Equation (\ref{eqn:sbpar}) is the same for the two solutions $\Delta
u_{0,-,\pm}$ or for the two solutions $\Delta u_{0,+,\pm}$, but is not
the same between these two pairs.  Because it is only the amplitude of
the parallax vector that enters the lens mass and distance, degeneracies
in solutions that affect only the direction of $\bpi_\e$ are relatively
unimportant.

As pointed out by \citet{gouldhorne}, the presence of a planet can
resolve the second (amplitude) degeneracy.  If the planetary caustic
appears in both light curves then this can prove, for example, that the
source trajectory appeared on the same side of the lens for the two
observatories.  This turns out to be the situation here.

\begin{figure}
\plottwo{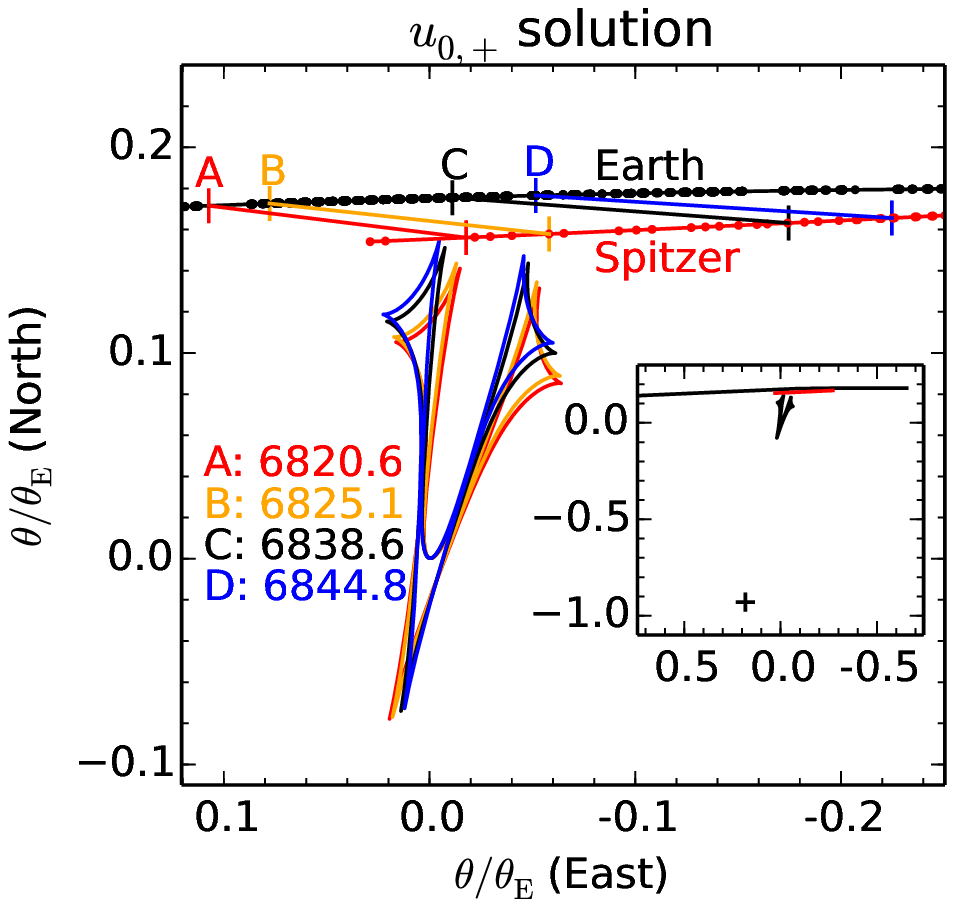}{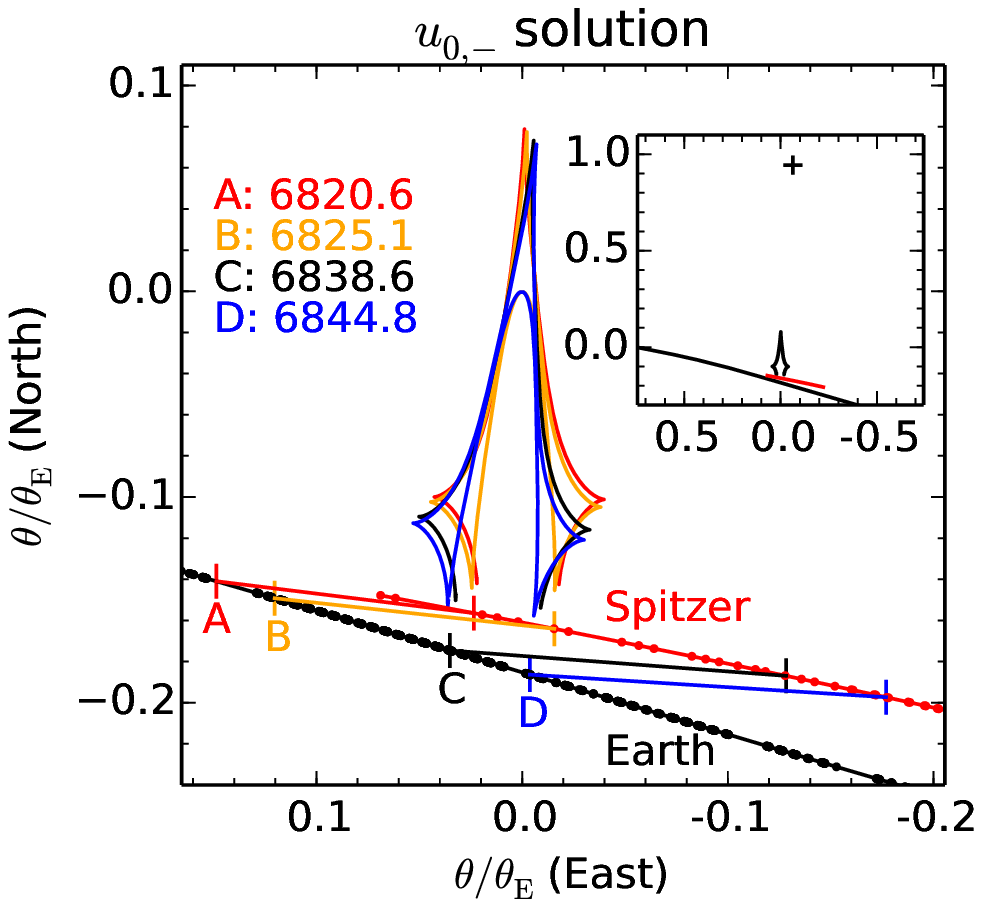}
\caption{Lens geometry for OGLE-2014-BLG-0124.  North is up and East is left.
The lens {\it position} is fixed, but with its orientation rotating
at $d\alpha/dt$ and the planet-star separation changing by $ds/dt$,
with the four epochs at which the source passes the two caustic ``prongs''
as seen from {\it Spitzer} and Earth shown in different colors.  The
source positions as seen from OGLE (black) and {\it Spitzer} (red) are shown
for each epoch of observation.  These trajectories deviate slightly
from rectilinear motion because of parallax effects of each observatory's
motion.  The line segments indicate common times at the two
observatories, which illustrate that the Earth-{\it Spitzer} projected
separation increases substantially over the 35 days of {\it Spitzer}
observations.  The left (right) panel shows the geometry of the $u_0>0$
($u_0<0$) solutions, which are very similar except for orientation
(see Tables~\ref{tab:ulens_so} and \ref{tab:phys_so}).  Planet location is
indicated by ``+'' symbols in insets.
}
\label{fig:caust}
\end{figure}

Nevertheless, the first degeneracy ($\pm u_0$) does persist.  The
geometries of the two solutions are illustrated in
Figure~\ref{fig:caust},  and the parameter values are listed in
Tables~\ref{tab:ulens_so} and \ref{tab:phys_so}. Note that the $u_0<0$
solution is disfavored by $\chi^2$, but is not completely excluded.

{\section{Two Tests of Earth-Orbit-Based Microlensing Parallax}
\label{sec:twotest}}

{\subsection{Fit to Ground Based Data of OGLE-2014-BLG-0124}
\label{sec:nonspitz}}

The Einstein timescale of this event was unusually long,
$t_\e=150\,$days. Such events very often yield parallax measurements,
particularly when the parallax is relatively large and the source is
relatively bright as in the present case.  It is therefore useful to
check the parallax measurement that can be made from just ground-based
data for two reasons. First, we would like to quantify the improvement
that is achieved by incorporating {\it Spitzer} data.  Second, we would
like check whether ground-based parallaxes (which rely on subtle
lightcurve effects that are potentially corrupted by systematics) agree
with a very robust independent determination.  In fact, of the dozens of
microlens parallax measurements that have been made (from a total of
$>10^4$ events), there has been only one completely rigorous test and
one other quite secure test (Section~\ref{sec:direct}).

\begin{figure}
\plotone{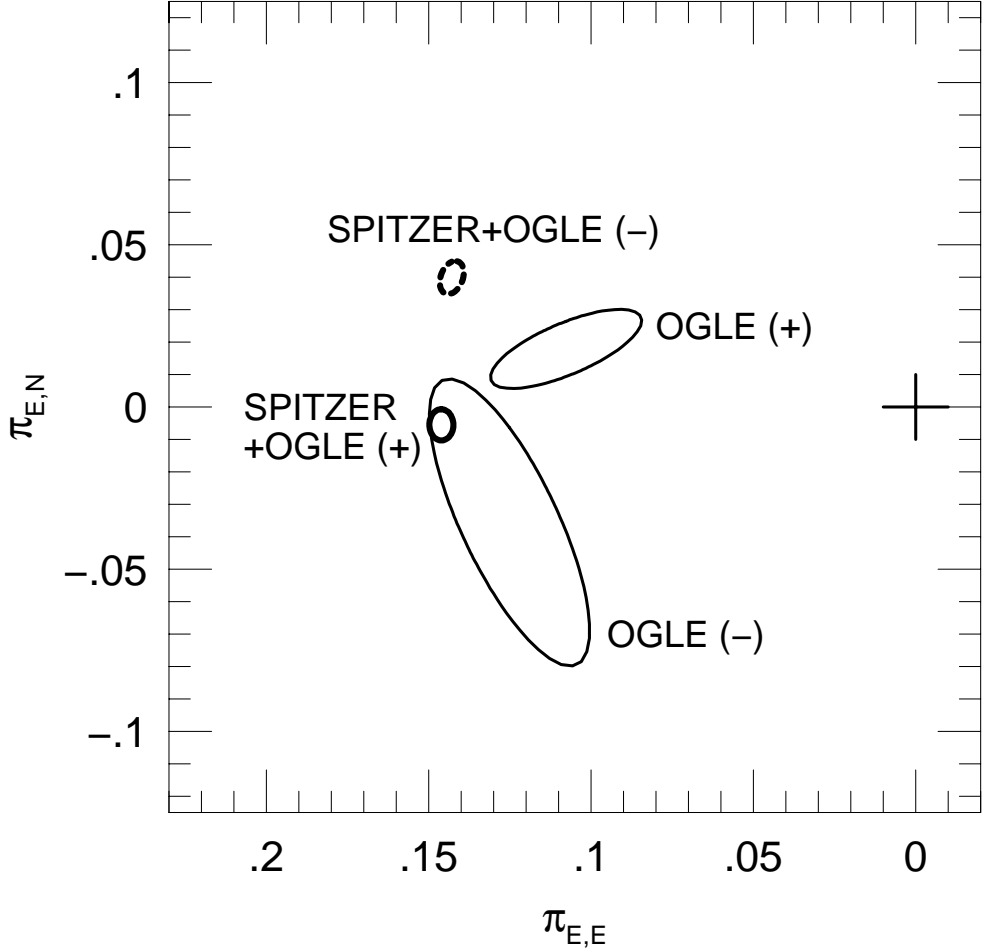}
\caption{Error contours ($\Delta\chi^2=1$) in the $\bpi_\e$ plane for two
solutions ($u_0>0$ and $u_0<0$) for each of two cases
(OGLE-only and {\it Spitzer}+OGLE), shown in standard and bold
curves, respectively.  For {\it Spitzer}+OGLE, the $u_0<0$ solution
is displayed as a dashed curve as a reminder that this solution
is disfavored but not formally excluded $(\Delta\chi^2=7)$.
By contrast, the OGLE-only solutions differ by $\Delta\chi^2=1$.
The $u_0>0$ and $u_0<$ solutions are indicated by (+) and (-), 
respectively.
}
\label{fig:ell3}
\end{figure}

We repeat the same procedures described in Sections~\ref{sec:anal} and
\ref{sec:phys} except that we exclude {\it Spitzer} data.  We report the
results in Tables~\ref{tab:ulens_o} and \ref{tab:phys_o}.
Figure~\ref{fig:ell3} compares the constraints on the parallax vector
from the OGLE data alone and the joint fit to the OGLE and {\it Spitzer}
data.

The first point is that with only OGLE data, the $u_0>0$ and $u_0<0$ 
solutions  are statistically indistinguishable (the degeneracy of
direction). For the dominant East component, these yield $\pi_{\e,E}=
0.108\pm 0.023$ and $0.125\pm 0.025$, i.e., 21\% and 20\% errors,
respectively. Since both solutions must be considered viable, we should
adopt $\bpi_\e=(-0.009,0.116)\pm (0.039,0.026)$ as the ``prediction'' of
the OGLE data.

Second, in contrast to many past parallax measurements, which are
typically much more constraining in the direction of Earth's
acceleration (East), this measurement has comparable errors in the North
and East directions.  This is undoubtedly due to the very long $t_\e$,
since analyses by \citet{gmb94}, \citet{smith03}, and \citet{gould04},
all show that so-called ``one-dimensional parallaxes'' explicitly arise
from the shortness of events relative to a year.

Third, the ground-based parallax measurements are off by 0.1 and 1.1
sigma respectively.  The probability for this level of discrepancy, 
assuming purely statistical errors, is $\sim 54\%$, i.e., quite
consistent.

Fourth, including {\it Spitzer} data improves the precision by a factor
of 7 in the East direction and a factor of 8 in the North direction.
This demonstrates the tremendous power of space-based parallaxes
relative to the ground, even for an event whose characteristics make it
especially favorable for ground-based measurement.

Finally, we note that while the parallax measurements with and without
{\it Spitzer} data are consistent at the $1\,\sigma$ level using the
OGLE-only error bar,  the derived lens mass and distance both show much
closer agreement relative to their statistical errors. This is because
the errors in both quantities are dominated by the errors in $\rho$
(through $\theta_\e$) and this quantity is poorly determined in the
present case.

\subsection{{A Second Direct Test: MACHO-LMC-5}
\label{sec:direct}}

{\it Spitzer} observations of OGLE-2014-BLG-0124 provide only the second
direct test of a microlens parallax measurement derived from so-called
``orbital parallax'', i.e., distortions in the lightcurve due to the
accelerated motion of Earth.  Such tests are quite important because
microlens parallaxes are derived from very subtle deviations in the
lightcurve, which could potentially be corrupted by -- or be even
entirely caused by -- instrumental systematics and/or real physical
processes unrelated to Earth's motion.

In the present case, we found that the accuracy of the ground based
measurement of ``orbital parallax'' (as judged by the comparison to the
much more precise Earth-{\it Spitzer} measurement) was nearly as good as
the relatively small formal errors of $\sigma(\bpi_\e)=(0.039,0.026)$.

\begin{figure}
\plotone{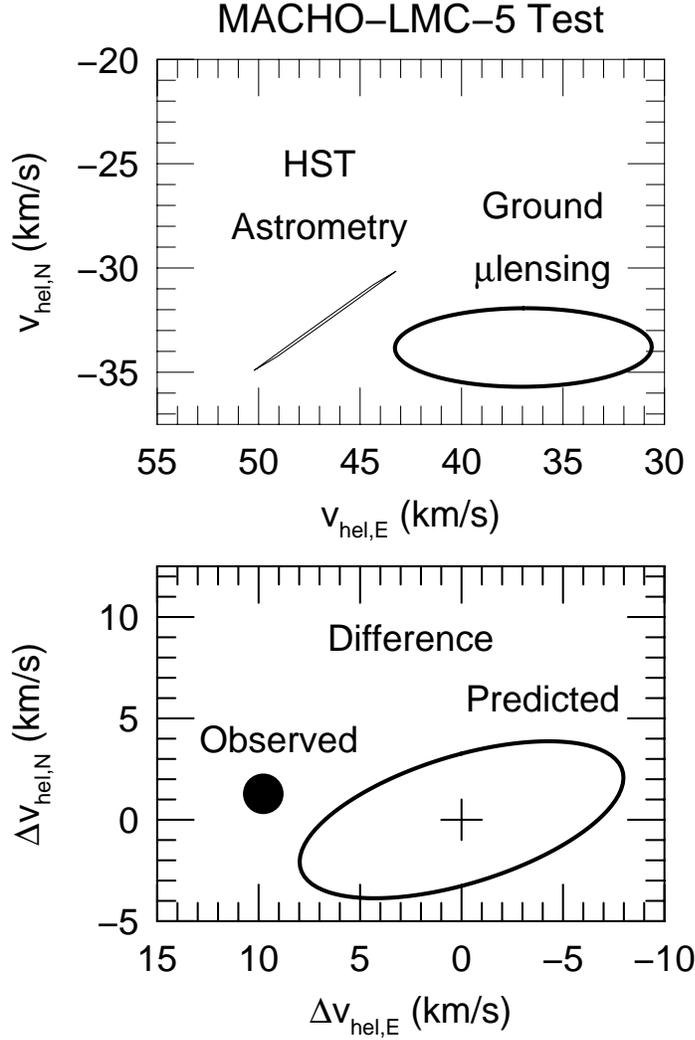}
\caption{Upper Panel: Comparison of projected velocity $\tilde \bv$ as 
determined from microlensing lightcurve and {\it HST} astrometry
for MACHO-LMC-5, which was discovered by the MACHO group in 1993.
Lower Panel: Predicted difference (zero with error ellipse) between these
two measurements compared to observed difference.  The $\Delta\chi^2=2.87$
(for 2 dof) implies consistency at the 24\% level.  MACHO-LMC-5 is the only 
ground-based parallax measurement (other than OGLE-2014-BLG-0124) for which
such a rigorous test is possible.  Both events pass this test.
}
\label{fig:ell2}
\end{figure}

The only other event for which data exist to directly test a microlens
parallax measurement is MACHO-LMC-5, which was one of the first
microlensing events ever observed.  In fact, although these data permit
a full two-dimensional test of the parallax measurement, all three
papers that addressed this issue \citep{alcock01,gould04,gould04b}
considered only a one-dimensional test, namely, a comparison of the
direction of $\bpi_{\e,\hel}$ measured from the microlensing light curve
with the direction of $\bmu_\hel$ measured from  {\it Hubble Space
Telescope (HST)} astrometry.  What makes a two-dimensional test possible
is the astrometric measurement of $\pi_\rel$ (in additional to
$\bmu_\hel$), which was available to \citet{gould04b} (but not the other
two papers) via the work of \citet{drake04}.  Then one can directly
compare
\begin{equation}
\tilde \bv_{\hel,hst} = \bmu_\hel{\au\over\pi_\rel};
\qquad
\tilde \bv_{\hel,\mu\rm lens} = {\bpi_{\e,\geo}\au\over \pi_\e^2 t_\e} +\bv_{\oplus,\perp}.
\label{eqn:hst}
\end{equation}
We adopt the data set finally assembled by \citet{gould04b} and find
\begin{equation}
\tilde \bv_{\hel,hst} = [(-32.5,46.73)\,\kms]\times (1\pm 0.10)
\qquad
\tilde \bv_{\hel,\mu\rm lens} = (-33.8 \pm 6.3,37.0\pm1.9)\,\kms.
\label{eqn:hst2}
\end{equation}
The form of the first equation reflects that the errors are almost
perfectly correlated  (correlation coefficients $\rho=0.9997$) so that
the errors can affect the magnitude but not the direction of this
vector. By contrast, the errors in the second equation are almost
perfectly independent ($\rho= 0.015$). The $\Delta\chi^2=1$ error
ellipses are shown in Figure~\ref{fig:ell2}.  In the lower panel we show
the error ellipse predicted for the difference of the two measurements
(based on the sum of the covariance matrices) compared to the actual
difference in Equation~(\ref{eqn:hst2}). This yields $\chi^2=2.87$ for
two degrees of freedom, which has a probability $\exp(-\chi^2/2)=24\%$,
i.e., quite consistent.

In addition to this direct test, there is one previous indirect test.
For the case of the two-planet system OGLE-2006-BLG-109Lb,c, the mass
and distance derived from a combination of microlens parallax  and
finite source effects were $M=\theta_\e/\kappa \pi_\e = 0.51\pm
0.05\,M_\odot$ and $D_L=1.49\pm 0.12\,\kpc$. \citep{ob06109,ob06109b}. 
These predict a dereddened source flux of  $H_0 = M_H +
5\log(D_L/10\,{\rm pc})= 5.94+10.87=16.81$. From high-resolution Keck
imaging, \citet{ob06109b} found $H=17.09\pm 0.20$. They estimated an
extinction of $A_H=0.3\pm 0.2$.  Hence, the two estimates of $H$ differ
by $\Delta H =0.01\pm 0.28$, not accounting for intrinsic dispersion in
$H$ as a function of mass.

\section{{Discussion}
\label{sec:discuss}}

The projected velocity $\tilde \bv$ is both the most precisely and most
robustly measured physical parameter, but it is also the most puzzling. 
Recall from Section~\ref{sec:heuristic} that $\tilde \bv_\hel =
(0,105)\,\kms$ can be derived from direct inspection of the lightcurve,
values that are confirmed and measured to a precision of $3\,\kms$ by
the lightcurve analysis as summarized in Table~\ref{tab:phys_so}.  

From the magnitude $\tilde v_\hel \sim 100\,\kms$, one would conclude
that the lens is most likely at intermediate distance in the disk.  This
is because 
\begin{equation}
\mu = {\tilde v\over \au}\pi_\rel.
\label{eqn:vtilemu}
\end{equation}
Hence, for stars within 1--2 kpc of the Sun (so $\pi_\rel\sim\pi_l$), we
have $\tilde v_\hel \sim v_{\perp,\rel}$, i.e., the transverse velocity
of the star in the frame of the Sun.  Since very few stars are moving at
$\sim 100\,\kms$, it is unlikely that a nearby star would have $\tilde
v_\hel \sim 100\,\kms$.  By the same token, bulge lenses have
$\pi_\rel\la 0.03$, meaning that this projected velocity measurement
would correspond to $\mu_\geo \la 0.5\,\masyr$.  Since typical values
for bulge lenses are $\mu_\geo\sim 4\,\masyr$ and since the probability
of slow lenses scales $\propto \mu_\geo^2$, bulge lenses with this
projected velocity are also unlikely.  Hence, the estimate $D_l = 4.1\pm
0.6\,\kpc$ from Table~\ref{tab:phys_so} seems at first sight quite
consistent with these general arguments.

The problem is that the direction of $\tilde \bv_\hel$, almost due East,
is quite unexpected for disk lenses at intermediate distance.  The fact
that the Sun and the lens both partake of the Galaxy's flat rotation
curve, while the bulge sources have roughly isotropic proper motions
implies that the mean heliocentric relative proper motion should be 
$\langle \bmu_\hel\rangle = \bmu_{\rm SgrA*}=(5.5,3.2)\,\masyr$.  
Hence, for an assumed distance of $D_l = 4.1\,\kpc$ ($\pi_\rel =
0.12\,$mas), there is an offset
\begin{equation}
\Delta\bmu_\hel = \bmu_\hel - \langle \bmu_\hel\rangle = (-5.6,0.5)\,\masyr.
\label{eqn:dmu}
\end{equation}
While it is not impossible that the source star is responsible for this
motion (although it is relatively large considering that the
1-dimensional dispersion of bulge lenses is $\sigma_\mu\sim 3\,\masyr$)
or that there is some contribution from the peculiar motion of the lens
itself,  the problem is that this unusually large motion just happens to
be of just the right size and direction to push $\tilde v_N \sim 0$.  Of
course, the lens must be going in some direction, but East is a very
special direction in the problem because that is the projected direction
of the  Earth-{\it Spitzer} axis.  

One generic way to produce a spurious alignment between the inferred
direction of lens motion and the Earth-{\it Spitzer} axis is to
introduce ``noise'' in the sparse epochs of the early {\it Spitzer}
lightcurve. We do not expect instrumental noise at this level and do not see any
evidence of it in the late {\it  Spitzer} light curve.  However, one way
to introduce astrophysical ``noise'' would be to assume that the true
direction of motion was very different and that  the planetary deviation
seen in the {\it Spitzer} lightcurve was from a second unrelated planet.
 This would require some fine-tuning because fitting even 4-5 deviated
points to an already-determined lens geometry is not trivial.  However,
there is a stronger argument against this scenario: the ground-based
data by themselves predict the same general trajectory (albeit with
seven times larger errors), so that even without having seen the {\it
Spitzer} data, one would predict that {\it Spitzer} would see deviations
due to the ground-observed planet at approximately this epoch.

Hence, we conclude that while the alignment of $\tilde \bv_\hel$ with
the Earth-{\it Spitzer} axis is indeed a puzzling coincidence, there are
no candidate explanations for this other than chance alignment.

\section{{Conclusions}
\label{sec:conclude}}

OGLE-2014-BLG-0124 is the first planetary microlensing event with a
space-based measurement of the vector microlens parallax $\bpi_\e$.
Combining $\pi_\e$ and $\theta_\e$ provides a means to precisely measure
masses of the host star and planet in microlensing events. In most
planetary microlensing events, $\pi_\e$ is the limiting factor in
obtaining a direct measurement of the planet's mass (but see
\citealt{zhu}). In this case, the combination of data from both OGLE and
{\it Spitzer} gives an error in  the amplitude of the parallax that is
only 2.5\%, implying that it contributes negligibly to the uncertainty
in the host mass $M=\theta_\e/\kappa\pi_\e = 0.71\pm 0.22\,M_\odot$. 
Rather, in contrast to the great majority of planetary microlensing
events discovered to date, this uncertainty is dominated by the error in
$\theta_\e$. That is, whereas most current planetary events have caustic
crossings that yield a precise measurement of $\rho=\theta_*/\theta_\e$,
so that the fractional error in $\theta_\e$ is just that of $\theta_*$ 
(typically $\sim 7\%$),  OGLE-2014-BLG-0124 did not undergo caustic
crossings. Rather, there is an upper limit on $\rho$ because if it were
too big, the source would have approached close enough to a cusp to give
rise to detectable effects, and a lower limit because small $\rho$
implies large $\theta_\e=\theta_*/\rho$ and thus large mass and large
lens-source relative parallax $\pi_\rel = \theta_\e\pi_\e$.  The
combination would make the lens bright enough to be seen for 
$\rho<6.5\times 10^{-4}$.  Hence the mass of the planet is  $m=0.51\pm
0.16\,M_{\rm jup}$ and its projected separation is  $a_\perp=3.1\pm
0.5\,\au$.  It lies at a distance $D_L = 4.1\pm 0.6\,\kpc$ from the Sun.

The high precision of the Earth-{\it Spitzer} microlens  parallax allows
the first rigorous test of a ground-based $\bpi_\e$ measurement from
OGLE-only data, which yielded  a 22\% measurement of $\pi_\e$.  The {\it
Spitzer} data show that this measurement is correct to within
$1.1\,\sigma$.  We use archival data to construct a second test using
purely astrometric {\it HST} data  to confirm the two-dimensional vector
projected velocity $\tilde\bv$ for MACHO-LMC-5 that was derived from the
microlensing data.   These tests show that ground-based microlensing
parallaxes are reliable within their stated errors in the relatively
rare cases that they can be measured.

\acknowledgments

The OGLE project has received funding from the European Research Council
under the European Community's Seventh Framework Programme
(FP7/2007-2013) / ERC grant agreement no. 246678 to AU.  Work by JCY was
performed under contract with the California Institute of Technology
(Caltech)/Jet Propulsion Laboratory (JPL) funded by NASA through the
Sagan Fellowship Program executed by the NASA Exoplanet Science
Institute.  AG was supported by NSF grant AST 1103471 and NASA grant
NNX12AB99G. Work by CH was supported by the Creative Research Initiative
Program (2009-0081561) of the National Research Foundation of Korea.  
This work is based in part on observations made with the Spitzer Space
Telescope, which is operated by the Jet Propulsion Laboratory,
California Institute of Technology under a contract with NASA.

\begin{table}                                                                  
\caption{\label{tab:ulens_so}                                                   
\sc  $\mu$lens Parameters (Spitzer+OGLE)}                                                     
\vskip 1em                                                                     
\begin{tabular}{@{\extracolsep{0pt}}llrr}                                      
\hline                                                                         
\hline                                                                         
Parameter & Unit & $u_0>0$ & $u_0<0$ \\                                        
\hline \hline                                                                  
$\chi^2/$dof        &                &  6664&  6671\\
                    &                & / 6769& / 6769\\
 \hline
$t_0 - 6800$        &day             &36.176&36.140\\
                    &                & 0.039& 0.040\\
 \hline
$u_0$               &                &  0.1749& -0.1778\\
                    &                &  0.0039&  0.0032\\
 \hline
$t_{\rm E}$         &day             &152.1&151.8\\
                    &                &  2.9&  2.4\\
 \hline
$s$                 &                &  0.9443&  0.9429\\
                    &                &  0.0030&  0.0023\\
 \hline
$q$                 &$10^{-3}$       & 0.694& 0.705\\
                    &                & 0.046& 0.038\\
 \hline
$\alpha$            &deg             & 78.216& -78.307\\
                    &                &  0.090&  0.100\\
 \hline
$\rho$              &$10^{-3}$       &  1.25&  1.37\\
                    &                &  0.38&  0.42\\
 \hline
$\pi_{\rm E,N}$     &                & -0.0055&  0.0399\\
                    &                &  0.0048&  0.0052\\
 \hline
$\pi_{\rm E,E}$     &                &  0.1461&  0.1430\\
                    &                &  0.0037&  0.0037\\
 \hline
$\gamma_\parallel$  &${\rm yr}^{-1}$ &-0.115&-0.119\\
                    &                & 0.017& 0.016\\
 \hline
$\gamma_\perp$      &${\rm yr}^{-1}$ &  0.77& -0.97\\
                    &                &  0.53&  0.45\\
 \hline
 \hline
\end{tabular}                                                                  
\end{table}                                                                    

\begin{table}                                                                  
\caption{\label{tab:phys_so}                                            
\sc Physical Parameters (Spitzer+OGLE)}                                    
\vskip 1em                                                                     
\begin{tabular}{@{\extracolsep{0pt}}llcc}                                      
\hline                                                                         
\hline                                                                         
Parameter & Unit & $u_0>0$ & $u_0<0$ \\                                        
\hline \hline                                                                  
$M_{\rm host}$                          &$M_\odot$    &  0.71&  0.65\\
                                        &             &  0.22&  0.22\\
 \hline
$M_{\rm planet}$                        &$M_{\rm jup}$&  0.51&  0.47\\
                                        &             &  0.16&  0.15\\
 \hline
Distance                                &kpc          &  4.10&  4.23\\
                                        &             &  0.59&  0.59\\
 \hline
$a_{\perp}$                             &AU           &  3.11&  2.97\\
                                        &             &  0.49&  0.51\\
 \hline
$\tilde v_{\rm N,hel}$                  &km/s         &  -3.0&  20.6\\
                                        &             &   2.6&   2.9\\
 \hline
$\tilde v_{\rm E,hel}$                  &km/s         & 107.0& 103.2\\
                                        &             &   2.8&   2.3\\
 \hline
$\mu_{\rm hel}$                         &mas/yr       &  2.77&  2.56\\
                                        &             &  0.86&  0.83\\
 \hline
$\beta=(E_{\rm kin}/E_{\rm pot})_\perp$ &             &  0.47&  0.57\\
                                        &             &  0.29&  0.30\\
 \hline
$\theta_{\rm E}$                        &mas          &  0.84&  0.78\\
                                        &             &  0.26&  0.25\\
 \hline
 \hline
\end{tabular}                                                                  
\end{table}                                                                    

\begin{table}                                                                  
\caption{\label{tab:ulens_o}    
\sc  $\mu$lens Parameters (OGLE-only)}        
\vskip 1em                                                                     
\begin{tabular}{@{\extracolsep{0pt}}llrr}                                      
\hline                                                                         
\hline                                                                         
Parameter & Unit & $u_0>0$ & $u_0<0$ \\                                        
\hline \hline                                                                  
$\chi^2/$dof        &                &  6621&  6622\\
                    &                & / 6732& / 6732\\
 \hline
$t_0 - 6800$        &day             &36.170&36.182\\
                    &                & 0.051& 0.054\\
 \hline
$u_0$               &                &  0.2099& -0.1964\\
                    &                &  0.0197&  0.0201\\
 \hline
$t_{\rm E}$         &day             &131.1&140.7\\
                    &                &  9.7& 13.1\\
 \hline
$s$                 &                &  0.9260&  0.9366\\
                    &                &  0.0096&  0.0097\\
 \hline
$q$                 &$10^{-3}$       & 0.752& 0.696\\
                    &                & 0.092& 0.086\\
 \hline
$\alpha$            &deg             &78.514&-78.566\\
                    &                &  0.183&  0.168\\
 \hline
$\rho$              &$10^{-3}$       &  1.60&  1.36\\
                    &                &  0.46&  0.44\\
 \hline
$\pi_{\rm E,N}$     &                &  0.0179& -0.0356\\
                    &                &  0.0122&  0.0443\\
 \hline
$\pi_{\rm E,E}$     &                &  0.1077&  0.1251\\
                    &                &  0.0233&  0.0247\\
 \hline
$\gamma_\parallel$  &${\rm yr}^{-1}$ &-0.148&-0.138\\
                    &                & 0.023& 0.023\\
 \hline
$\gamma_\perp$      &${\rm yr}^{-1}$ &  0.62& -0.52\\
                    &                &  0.68&  0.67\\
 \hline
 \hline
\end{tabular}                                                                  
\end{table}                                                                    

\begin{table}                                                                  
\caption{\label{tab:phys_o}                                                    
\sc Physical Parameters (OGLE-only)}                                           
\vskip 1em                                                                     
\begin{tabular}{@{\extracolsep{0pt}}llcc}                                      
\hline                                                                         
\hline                                                                         
Parameter & Unit & $u_0>0$ & $u_0<0$ \\                                        
\hline \hline                                                                  
$M_{\rm host}$                          &$M_\odot$    &  0.81&  0.74\\
                                        &             &  0.20&  0.21\\
 \hline
$M_{\rm planet}$                        &$M_{\rm jup}$&  0.63&  0.53\\
                                        &             &  0.18&  0.16\\
 \hline
Distance                                &kpc          &  4.92&  4.25\\
                                        &             &  0.69&  0.72\\
 \hline
$a_{\perp}$                             &AU           &  3.16&  3.13\\
                                        &             &  0.46&  0.47\\
 \hline
$\tilde v_{\rm N,hel}$                  &km/s         &  28.9& -28.0\\
                                        &             &  38.6&  30.5\\
 \hline
$\tilde v_{\rm E,hel}$                  &km/s         & 149.6& 111.5\\
                                        &             &  27.5&  17.2\\
 \hline
$\mu_{\rm hel}$                         &mas/yr       &  2.48&  2.81\\
                                        &             &  0.64&  0.86\\
 \hline
$\beta=(E_{\rm kin}/E_{\rm pot})_\perp$ &             &  0.40&  0.38\\
                                        &             &  0.31&  0.30\\
 \hline
$\theta_{\rm E}$                        &mas          &  0.72&  0.83\\
                                        &             &  0.19&  0.26\\
 \hline
 \hline
\end{tabular}                                                                  
\end{table}                                                                    

\end{document}